\documentclass[conference]{IEEEtran}
\IEEEoverridecommandlockouts

\usepackage{cite}
\usepackage{amsmath,amssymb,amsfonts}
\usepackage{algorithmic}
\usepackage{graphicx}
\usepackage{textcomp}
\usepackage{xcolor}
\usepackage[nolist,nohyperlinks]{acronym}
\usepackage{booktabs}
\usepackage[disable]{todonotes}
\usepackage{url}
\usepackage{subfig}

\newcommand{\N}{\ensuremath{\mathbb{N}}}
\newcommand{\R}{\ensuremath{\mathbb{R}}}  

\newcommand{\M}{\ensuremath{\mathbb{M}}} 
\newcommand{\C}{\ensuremath{\mathbb{C}}}

\def\BibTeX{{\rm B\kern-.05em{\sc i\kern-.025em b}\kern-.08em
    T\kern-.1667em\lower.7ex\hbox{E}\kern-.125emX}}
\begin{document}
\begin{acronym}[XXXXXXXX]
    \acro{ADC}{analog to digital converter}
    \acro{DAC}{digital to analog converter}
    \acro{DNN}{deep neural network}
	\acro{FFT}{fast Fourier transform}
    \acro{RF}{radio frequency}
    \acro{SDR}{software defined radio}
    \acro{SFE}{Synchronization Feature Estimator}
    \acro{SER}{symbol error rate}
    \acro{SNR}{signal-to-noise ratio}
    \acro{TF}{TensorFlow}
\end{acronym}

\title{A Deep Learning Wireless Transceiver with Fully Learned Modulation and Synchronization
{}
\thanks{}
}

\author{\IEEEauthorblockN{Johannes Schmitz, Caspar von Lengerke, Nikita Airee,\\ Arash Behboodi, Rudolf Mathar}
\IEEEauthorblockA{\textit{Institute for Theoretical Information Technology} \\
\textit{RWTH Aachen University} \\
    52074 Aachen, Germany
    }
}

\maketitle

\begin{abstract}
In this paper, we present a deep learning based wireless transceiver. We describe in detail the corresponding artificial neural network architecture, the training process, and report on excessive
over-the-air measurement results.
We employ the end-to-end training approach with an autoencoder model that includes a channel model in the middle layers as  previously proposed in the literature.
In contrast to other state-of-the-art results, our architecture supports learning time synchronization without any manually designed signal processing operations.
Moreover, the neural transceiver has been tested over the air with an implementation in software defined radio.
Our experimental results for the implemented single antenna system demonstrate a raw bit-rate of 0.5 million bits per second.
This exceeds results from comparable systems presented in the literature and suggests the feasibility of high throughput deep learning transceivers.
\end{abstract}

\section{Introduction}
Deep learning techniques had a tremendous impact on numerous research areas over the last decade.
While groundbreaking results in computer vision have been reported already at the beginning of the decade, deep learning methods for communication systems have gained attraction only recently. 
Although many works already proposed machine learning solutions to communication system problems, for example, modulation detection in \cite{oshea2016convolutional}, the first end-to-end deep learning communication systems appeared in \cite{oshea2017introduction} and \cite{dorner2018dl}.
Previously the design of communication systems relied on information theoretic models, classical optimization techniques and signal processing algorithms that were optimal for certain systems.
These approaches, however, are limited to only mathematically tractable channel models and can in most cases only be applied to isolated components of the digital communication chain.
In contrast, the novel deep learning approach promises to overcome these limitations by applying a global end-to-end optimization to the problem at hand.
Learning based approaches do not need a tractable channel model and optimize the system globally.
A \ac{DNN} autoencoder was proposed in \cite{oshea2017introduction} as a first attempt to end-to-end learning of communication systems. The autoencoder structure, a multilayer neural network,  consists of an encoder-decoder pair that models the transmission of bits on the physical layer with raw data bits as the autoencoder's input and output.

A channel model is included in the middle layers, and the modulation and demodulation are implemented by the first and last layers of the autoencoder. After training, the first layers are separately used as the transmitter. The final layers constitute the receiver of the system. The middle layers, which model the effect of channels, are omitted after training. 
In order to adapt the communication system to arbitrary real-world channels lacking tractable mathematical models, some extensions of autoencoders have been proposed in \cite{aoudia2018end,ye2018channel,raj2018backpropagating}.

In this paper we model the effect of channel impairments through customized middle layers of the autoencoder.
For achieving practical applicability it is assumed that the autoencoder incorporates a digital complex baseband model, where signal samples generated by the transmitter \ac{DNN} can directly be fed to an \ac{RF}-frontend for up-conversion.
Down-converted signal samples from the receiver \ac{RF}-frontend can be directly fed into the receiver \ac{DNN} without any manual pre-processing.

For such a complete end-to-end deep learning signal processing approach the problem of synchronization arises. 
The authors of \cite{oshea2016rtn} and \cite{dorner2018dl} discuss radio transformer networks for this purpose. Signal parameters are estimated with a separate \acp{DNN} and the signal is then processed by manually programmed operations. The authors in \cite{zhao2018deep} and \cite{felix2018ofdm} propose to learn merely the receiver side of the system and afterwards apply manual signal processing to support synchronization. 

We present instead a new full end-to-end approach where synchronization is completely learned as part of a single \ac{DNN} structure. This is achieved by pilot symbols with a learned waveform and by feeding longer signals to the decoder.
This paper is structured as follows.
In Sec.~\ref{sec:architecture} we describe the architecture of the autoencoder. The training process is described explicitly in Sec.~\ref{sec:channel_model}.
The details of the \ac{SDR} implementation are elaborated in Sec.~\ref{sec:sdr}.
The paper concludes with the presentation and discussion of the test results in Sec.~\ref{sec:results}.

\section{A Deep Neural Network Transceiver}
\label{sec:architecture}
In this section we describe the structure of the \ac{DNN} autoencoder, dimensions and parameters are given in Tab.~\ref{tab:dnn_layers}.
The transmitter is represented by the encoder and the receiver by the decoder, both connected by the channel. The autoencoder operates in discrete passes $i \in \N$.
In each pass a source symbol $s=(b_1,\ldots,b_k)\in\{0,1\}^k$ of $k$ bits
is fed into the transmitter.
The outputs of the transmitter are $n$ complex samples comprised in a vector $\mathbf{x}=(x_1,\ldots,x_n) \in \C^{n}$.

These samples are transformed by a \ac{DAC} to a baseband waveform and transmitted after up-conversion. Down-conversion and sampling yields the noisy samples $\mathbf{y}=(y_1,\ldots,y_n) \in \C^{n}$ as the output of the channel. The decoder receives a sequence of $W$ output samples as input and then decides which original sequence was sent. The decoder input length $W$ is larger than $n$ to combat synchronization errors. We will discuss the choice of $W$ below.

The autoencoder \textit{encodes} a sequence $s$ of $k$ bits into a sequence of $n$ complex samples and \textit{decodes} $W$ complex samples back into a sequence $\hat s$ of bits. There are $M = 2^k$ possible symbols $s$ and estimates $\hat{s}$ so that the autoencoder represents a classification network. We use categorical cross entropy as error function during training.

\begin{table}[bht]
\vspace{2mm}
\caption{Layout of AE-7/16}
\resizebox{\columnwidth}{!} {%
\begin{tabular}{lcc}
\toprule
Encoder:            & Parameters & Output Dimensions\\
\midrule
Input               & 0         & 1 (integer s $\in [0,127]$) \\
Embedding           & 16384     & 128   \\
Dense (ReLU)        & 16512     & 128 \\
Dense (ReLU)        & 4128      & 128 \\
Dense (Linear)      & 1056      & 32 \\
Normalization       & 0         & 32 \\
Real2Complex	    & 0         & 16 \\
\midrule
Channel:            &           & \\
\midrule
Serialize           & 0         & 80 \\
Time Shift          & 0         & 47 \\
Phase Noise	        & 0         & 47 \\
Gaussian Noise	    & 0         & 47 \\
Multiply	        & 0         & 47 \\
\midrule
Receiver:           &           & \\
\midrule
Complex2Real        & 0         & 94 \\
\midrule
\multicolumn{3}{l}{Synchronization Feature Estimator (SFE):}\\
\midrule
Reshape		  		& 0         & (47,2) \\
Convolution (ReLU)  & 896       & (45,128) \\
MaxPool             & 0         & (45,128) \\
Convolution (ReLU)  & 131136    & (30,64)\\
MaxPool             & 0         & (15,64) \\
Flatten             & 0         & 960 \\
Dense (ReLU)        & 492032    & 512 \\
Dense (ReLU)        & 5130      & 10\\
\midrule
Decoder:            &           & \\
\midrule
Concatenate		    & 0         & 104 \\
Dense (ReLU)        & 53760     & 512 \\
Dense (ReLU)        & 262656    & 512 \\
Dense (ReLU)        & 131328    & 256 \\
Dense (ReLU)        & 65792     & 256 \\
Dense (ReLU)        & 32896     & 128 \\
Dense (Softmax)     & 16512     & 128 \\
ArgMax			    & 0         & 1\\
\bottomrule \\
\end{tabular}}
\label{tab:dnn_layers}
\end{table}
\subsection{Encoder}
\label{sec:encoder}
Each input symbol $s$ is represented by an integer $1,2,\ldots,M=2^k$
and fed into an initial embedding layer.
\todo[inline]{following sentence: and implicit channel coding?}
\todo[inline]{Arash: After some though, I finally think that the term "modulation" is the best here. You are changing signal samples to transmit different symbols.}
The following two layers of $M$ neurons are fully connected and use the
ReLU activation function to implement modulation.
The encoder is supposed to learn a representation $\mathbf x \in \C^n$ for each $s$.
Mapping symbols $s$ to sequences of $n$ complex numbers is carried out by a
third dense layer without any activation function.
The width of this layer is $2n$, representing the real and imaginary part 
of $n$ complex samples. In the next layer all samples are normalized to fall into the unit circle of the complex plane, which is necessary to match the \ac{DAC} input specifications and essentially also due to the limited output power of the RF frontend. The final encoder layer then combines two real values into one complex number, thus generating the complex baseband samples.

\subsection{Channel model and training}
\label{sec:channel_model}
The present autoencoder transceiver cannot be trained without a channel model,
since backpropagation spans over all layers of the DNN. We use a channel model that incorporates additive white Gaussian noise (AWGN), phase shift, time shift and attenuation. All are modeled to be frequency independent, which is a simplification, sufficient for narrowband channels. The channel model hence extends over the standard memoryless channels to a more challenging setting with asynchronous communication.  

In real systems phase offset occurs due to asynchronous clocks or phase noise. Since the autoencoder cannot memorize the phase offset from previous transmissions, we model it as an independent, uniformly distributed random variable in each training step. This is reasonable as the phase offset changes only little for the duration of one symbol transmission.
Hence, all samples in the vector $\mathbf x$ generated by the encoder are rotated by a uniformly distributed phase 
 \begin{equation}
 \mathbf u = e^{-j\varphi} \cdot \mathbf x.
 \end{equation}
This operation is implemented as a layer of the channel model, where $\varphi$ is provided as an additional input to the layer. The assumptions above are harder than a real channel with correlation over time.


Attenuation of signals is modeled by multiplying each sample by a uniformly distributed random factor $a \in [a_{\text{min}},1]$ as 
\begin{equation}
\mathbf v = a\cdot\mathbf u.
\end{equation}

At the next step, the random noise is added to the symbols. Independent random noise variables are added to each complex sample
\begin{equation}
\mathbf y=\mathbf v+\mathbf r.
\end{equation}
A small value of \ac{SNR} is chosen, as the autoencoder does not generalize well for \ac{SNR} less than the training \ac{SNR}. We observed, moreover, that the autoencoder does not converge to a suitable state if the \ac{SNR} is too small. The autoencoder, however, converges during training and generalizes well for moderately larger \ac{SNR} values. Note that the noise is added to each complex sample. Therefore the \ac{SNR} is, in this sense, defined by $E_{\text{sample}}/N_0$. We can convert the \ac{SNR} per sample to the \ac{SNR} per bit, denoted by  $E_{\text{b}}/N_0$ using
\[
 E_{\text{b}}/N_0=\frac{k}n E_{\text{sample}}/N_0.
\]
The \ac{SNR} per bit is used to evaluate the performance independent of the number of complex samples $n$.

Finally, synchronization effects are introduced. Since the decoder has no memory, synchronization is a difficult problem for DNN based transceivers. In practice, synchronization errors occur gradually due to slightly asynchronous clocks at the transmitter and the receiver. To overcome this problem, in this work, the receiver takes account of a larger window of symbols that includes more than one baseband symbol. The purpose is that the receiver always has available the full set of samples for one of the data symbols within each window.

\begin{figure}
  \centering\includegraphics[width=0.95\columnwidth]{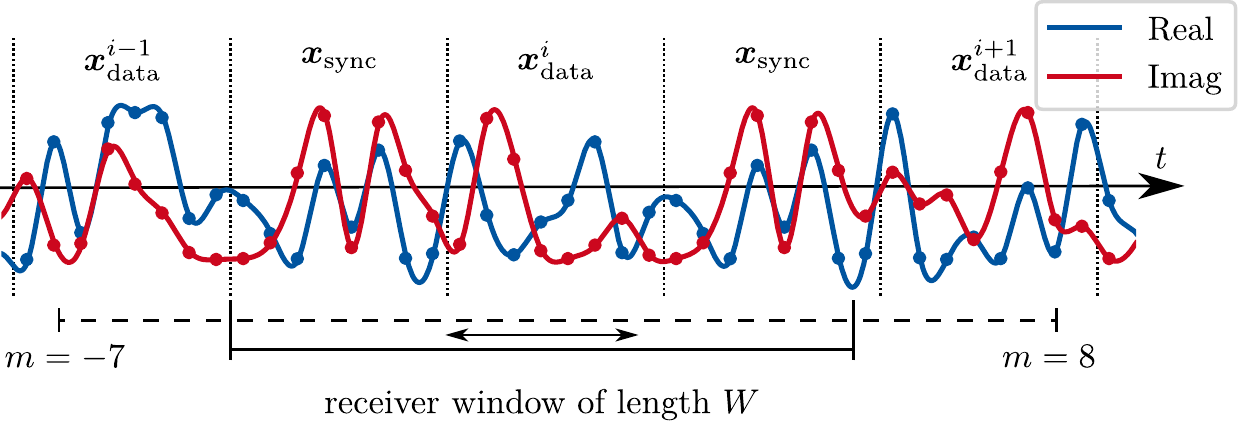}
  \caption{Visualizing the time shift for the channel and receiver of autoencoder AE-8/8 for a shift m = 0 samples}
  \label{fig:sliding_window}
\end{figure}

\begin{figure}
  \includegraphics[width=\columnwidth]{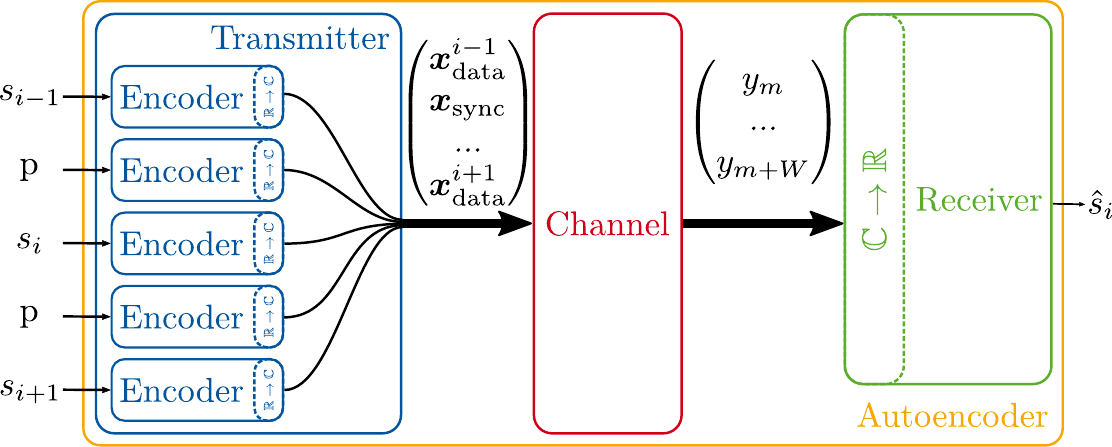}
  \caption{Autoencoder Layout during Training}
  \label{fig:autoencoder_shift}
\end{figure}

To introduce the time offset in the channel model, multiple baseband symbols are first considered. We use pilot symbols to facilitate synchronization. The pilot baseband samples are generated by the same encoder that generates the data baseband symbols. Hence, the encoder expects as input an alternating stream of pilot symbols $p$ and source symbols $s_i$. 
Fig.~\ref{fig:sliding_window} depicts an exemplary sequence of data and pilot channel symbols at the training step $i$ given by:
\[
(\mathbf x_{\text{data}}^{i-1}, \mathbf x_{\text{sync}}, \mathbf x_{\text{data}}^{i}, \mathbf x_{\text{sync}}, \mathbf x_{\text{data}}^{i+1})\in\C^{5n}.
\] 
The pilot symbol $p$ is chosen to be the integer $0\in\M$. The same number of bits is used for data and pilot symbols. At each training step, the channel symbols, for data and pilots, are generated by the encoder successively. We use the \textit{TimeDistributed} model of Keras to create 5 identical parallel encoders, as depicted in~Fig.~\ref{fig:autoencoder_shift}.
The weights of the parallel encoders are shared and updated jointly during training. The transmitter output consists of $5n$ consecutive complex samples during training. 
The channel impairments are added afterwards. 
First, a random phase shift $\varphi_i$ and a random attenuation $a_i$ are chosen. They act on each transmitted baseband sample $x \in \C$. White Gaussian noise $\mathbf{r}_i \in \R^W$, which satisfies SNR $E_{\text{sample}}/N_0$, is added at the end. The channel output at the training step $i$ is therefore given by
 \[
 \mathbf y^{(i)}=a_i e^{-j\phi_i}(\mathbf x_{\text{data}}^{i-1}, \mathbf x_{\text{sync}}, \mathbf x_{\text{data}}^{i}, \mathbf x_{\text{sync}}, \mathbf x_{\text{data}}^{i+1})+\mathbf r_i\in\C^{5n}.
 \]

The receiver uses a window of length $W = 3n-1$ samples, as to include at least one full data symbol regardless of the actual position of the window.
Without synchronization errors, this window is placed at the beginning of the first pilot and spans over the data symbol as in Fig.~\ref{fig:autoencoder_shift}. Synchronization errors introduce a shift of the window. At training step $i$, the window is shifted by an offset $m_i$ drawn from the uniform distribution over $\{-n+1,\dots,n\}$. 
Therefore, the input vector of complex baseband samples to the receiver in step $i$ is given by
\begin{equation}
\mathbf y^{(i)}[m_i] =  (y^{(i)}_{m_i},\dots,y^{(i)}_{m_i+W-1})\in\C^W.
\label{eq:windowreceiver}
\end{equation}

This approach provides a sufficient number of samples to the receiver to account for a set of possible time-shifts of the receiver window. Note that, in contrast to over the air deployment, we do not receive a continuous sequence of complex samples during training. Each time only $5n$ samples are generated over which the receiver chooses a window of $W$ samples. It should be noted that by introducing the synchronization symbol, the communication rate is halved. 

\subsection{Decoder}

The final step in the transceiver design is the decoder. It receives $2W$ real inputs
derived from  $W$ complex symbols. At training step $i$, the decoder shall correctly identify the source symbol $\mathbf s_i$ from the input of $W$ complex samples. 
It is trained to cope with synchronization errors and other channel impairments and,
furthermore, includes a particular entity to support the synchronization task called \ac{SFE}. \ac{SFE} extracts a set of features using convolutional layers (see Table \ref{tab:dnn_layers}). The decoder then uses these features with $2W$ received inputs to perform the final decoding in the next layers. \ac{SFE} is similar to  correlation filters in conventional transceivers which are used for synchronization. 

The final output of the decoder is a probability vector 
$\mathbf{\hat s}\in[0,1]^M$ with components $e^{vj}/\sum_\ell e^{v_\ell}$, where $v_j$, $j=1,\ldots,M$, denotes the real output of the final layer. The symbol with the highest probability is then chosen as the most likely input symbol.

\section{Software defined radio implementation}
\label{sec:sdr}
To test the trained transmitter and receiver \acp{DNN} in practice, we have implemented a custom signal processing block for the GNU Radio software defined radio (\ac{SDR}) \cite{gnuradio} framework. 
The Autoencoder is trained with \ac{TF} and the Keras code, written in Python. Thereafter, the DNNs are exported and loaded into the GNU Radio block, which is implemented in C++. This block is able to run trained \ac{TF} \cite{abadi2016tensorflow} models, i.e. perform the inference, in C++ \cite{schmitz2018grtensorflowcc}.
GNU Radio provides easy ways to interface with the \ac{RF} hardware frontends.
The autoencoder transceiver operates only in baseband. Hence, for radio transmission over the air, up- and down-conversion to and from the carrier frequency is performed by the \ac{RF} frontends. We used the 2.4GHz frequency band for transmission.
The C++ implementation of the system leads to higher throughput compared to earlier python implementations, e.g., in \cite{dorner2018dl}.

In this work, the signal-to-noise ratio of the AWGN is set to $E_{\text{sample}}/N_0 = 5\,dB$. As mentioned above, the autoencoder cannot be trained properly for too small \ac{SNR}, e.g., 0\,dB. 

By blockwise processing with one pilot per data block and a bandwidth of 1\,MHz, we achieve a throughput of 0.5\,Mbit/s on a machine with an Intel Core i7 940 CPU and NVIDIA GeForce GTX 1080 Ti GPU. Two USRP N210 \ac{RF}-frontends from Ettus were used for the experiments.

In the current setup, the ratio of pilot to data symbols is $1$. To improve the throughput, more data symbols per pilot symbol could be transmitted. The optimum ratio of pilot to data symbols will be determined in future experiments. 

\section{Experimental Results}
\label{sec:results}

The proposed system is evaluated by estimating the symbol error rate of transmissions over simulated and real channels. We compare the results with binary phase shift keying (BPSK) over perfect Gaussian channels. The BPSK transmission differs from our \ac{DNN} transceiver at least in two different aspects. First, no synchronization error is assumed for BPSK, and secondly, each bit is mapped to one complex channel sample. In contrast, our system considers synchronization errors and maps $k$ bits to $n$ complex samples. 

In our benchmark test series, different autoencoders are trained for varying parameters $k$ and $n$, denoted by AE-$k$/$n$. $k/n$ is the ratio between the source symbol bits and the number of complex baseband samples. For example, an AE-7/16 transmits 7 bits using 16 samples. After each data symbol of length $n$ a pilot symbol of the same length is inserted.
To evaluate the efficiency of \ac{SFE} 
the autoencoder named AE-8/8-2 is trained and tested without this unit.

\subsection{Over-the-air-transmission}
\begin{figure} 
	\centering
  \centering\includegraphics[width=0.9\columnwidth]{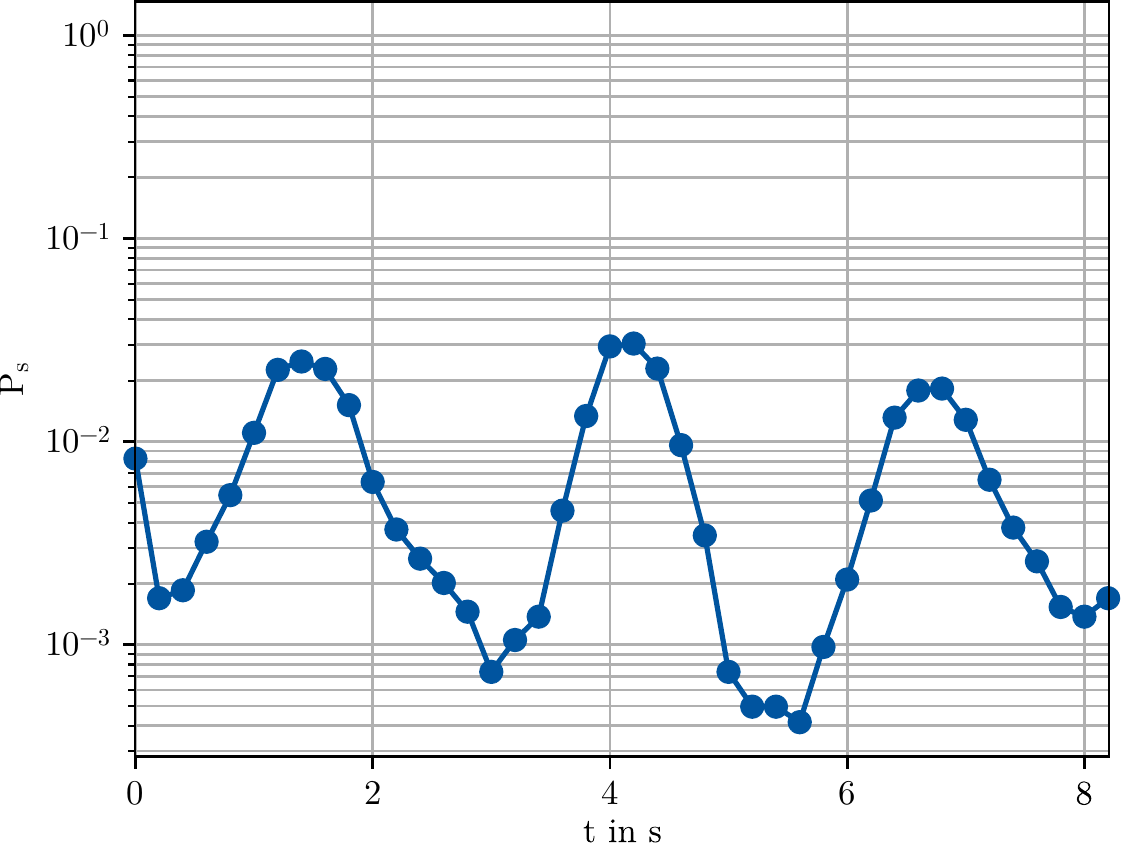}  
  \caption{SER of AE-8/8 tested over the air at amplitude 0.6156, windowed over 200ms for each data point, no overlap, for $1.25\cdot10^6$ symbols in total}
  \label{fig:ser_unsynched}
\end{figure}

To investigate transmission over a real channel, an AE-8/8 was tested over the air at a relative amplitude of 0.61. The error rate is calculated every 200\,ms, which corresponds to $200000/8= 25000$ symbols per evaluated time window.
The resulting SER is plotted over time in Fig. \ref{fig:ser_unsynched}.
Most notably, the error rate fluctuates periodically in intervals of about 2.5\,s.
This is attributed to the slightly different clock speeds of the two USRPs.
\todo[inline]{Caspar: Something is broken here now 32ppm is too much and the sentence is not clear, B210 has 1ppm}
The error rate changes, as drifting time offsets make it more difficult for the autoencoder to decode certain received sample windows.
The minimum at 5.5\,s corresponds to an error rate of $4\cdot10^{-3}$, which means a single error during a period of 200\,ms.

\begin{figure} 
	\centering
  \includegraphics[width=1.0\columnwidth]{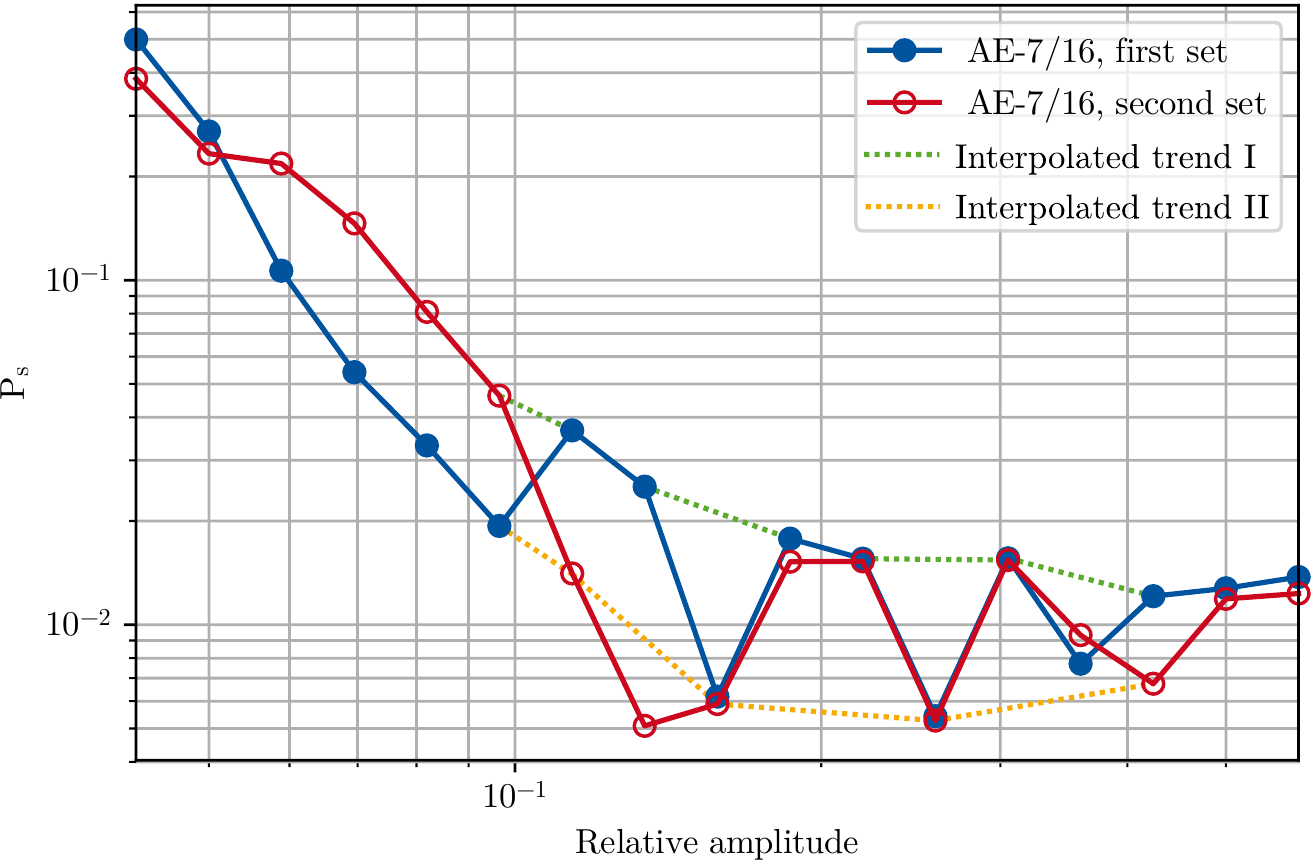}  
  \caption{SER of AE-7/16 tested over the air in two experiments for 3.4$\cdot10^6$ symbols each. The dotted lines emphasize patterns observed over the two sets.}
  \label{fig:ser_ota_716_unsyced}
\end{figure}

Next, after training AE-7/16 was tested over the air in two independent experiments for different relative amplitudes at the transmitter, each with $N_{\text{test}}=3.4\cdot10^6$ test symbols.
The resulting error rates are shown in Fig.~\ref{fig:ser_ota_716_unsyced}.
For a real radio transmission, the receiver \ac{SNR} cannot be measured precisely. Hence, the SER is plotted over the the transmit amplitude as $x$-axis.

Both experiments show a decrease of errors for higher amplitudes, as expected.
A higher amplitude causes more energy for the signal and  reduces the effect of noise at the receiver.
Very low amplitudes correspond to very low values of attenuation $a$ in the channel model, so that the neural receiver decodes increasingly incorrectly for amplitudes below 0.005.
This effect was also observed over-the-air, thus reducing performance at low levels of relative amplitude.
Both experiments show a bottom floor of \ac{SER} at around 1\%.

Notably, the error rates do not decrease monotonically with increasing amplitude. 
However, the main trend is indicated
by dotted lines. 
The upper green trend curve converges to an error rate of about 2\%, the yellow dotted line refers to error rates smaller by approximately a factor of 3.
At higher relative amplitudes the error rates increase caused by 
imperfections of the transmitter.
The reason for the two observed distinctly different error rate curves 
remains unclear at that point and will be investigated in the future.

The observed error rates are significantly worse than the ones found when testing the trained autoencoder over the corresponding simulated channel, results are depicted in Fig.~\ref{fig:ser_compare_model}. 
This can be attributed to the channel model that only approximates reality, however, is used to train the network. A lot of potential seems to be in applying more accurate channel models or finding ways to train the autoencoder by training even over real channels as has been shown in \cite{goutay2018deep}.

\subsection{Comparison of the autoencoders}
\begin{figure}
	\centering
	\subfloat[Subfigure 1 list of figures text][Normalized per bit]{
			\includegraphics[width=1.0\columnwidth]{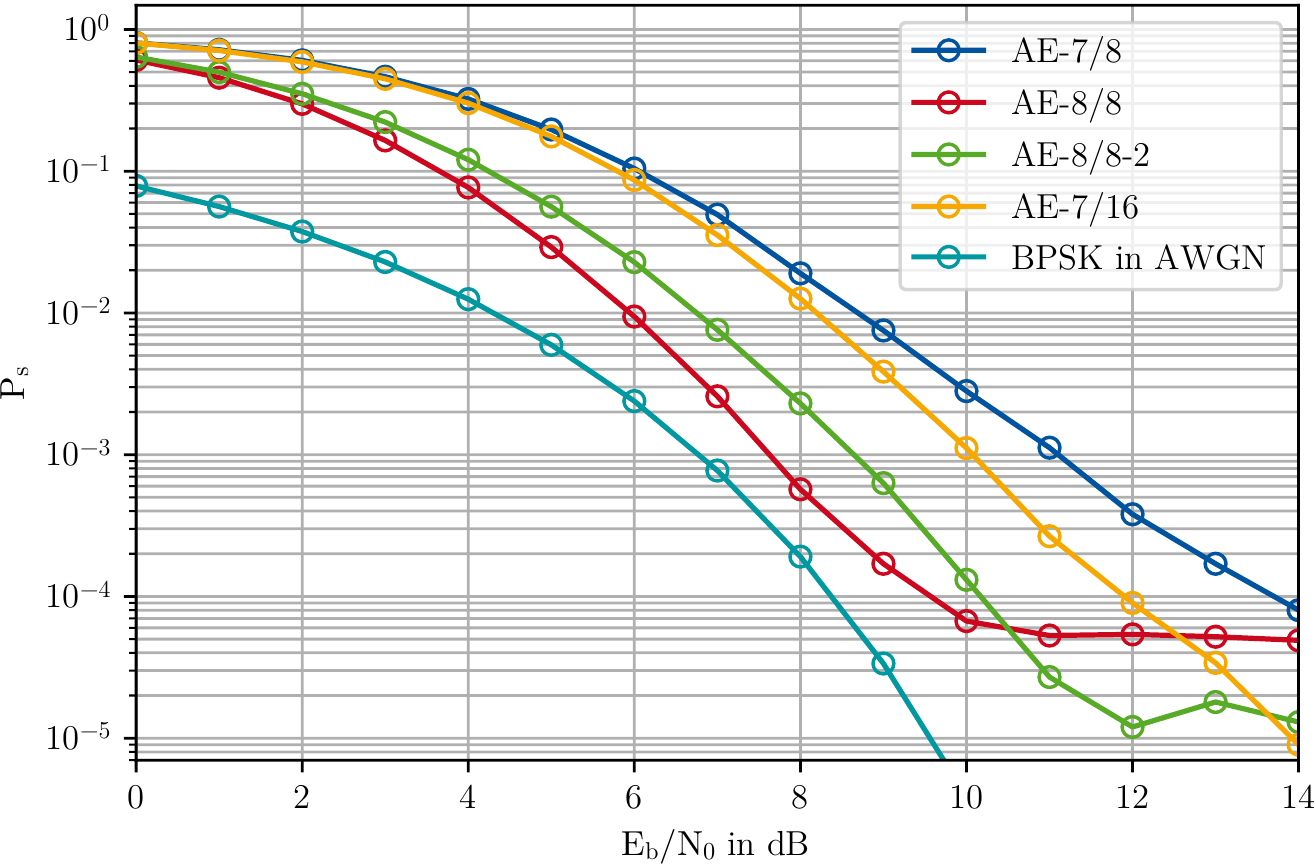}
		  \label{fig:ser_modelcompare}}
	\qquad
	\subfloat[Subfigure 2 list of figures text][Normalized per sample]{
		\includegraphics[width=1.0\columnwidth]{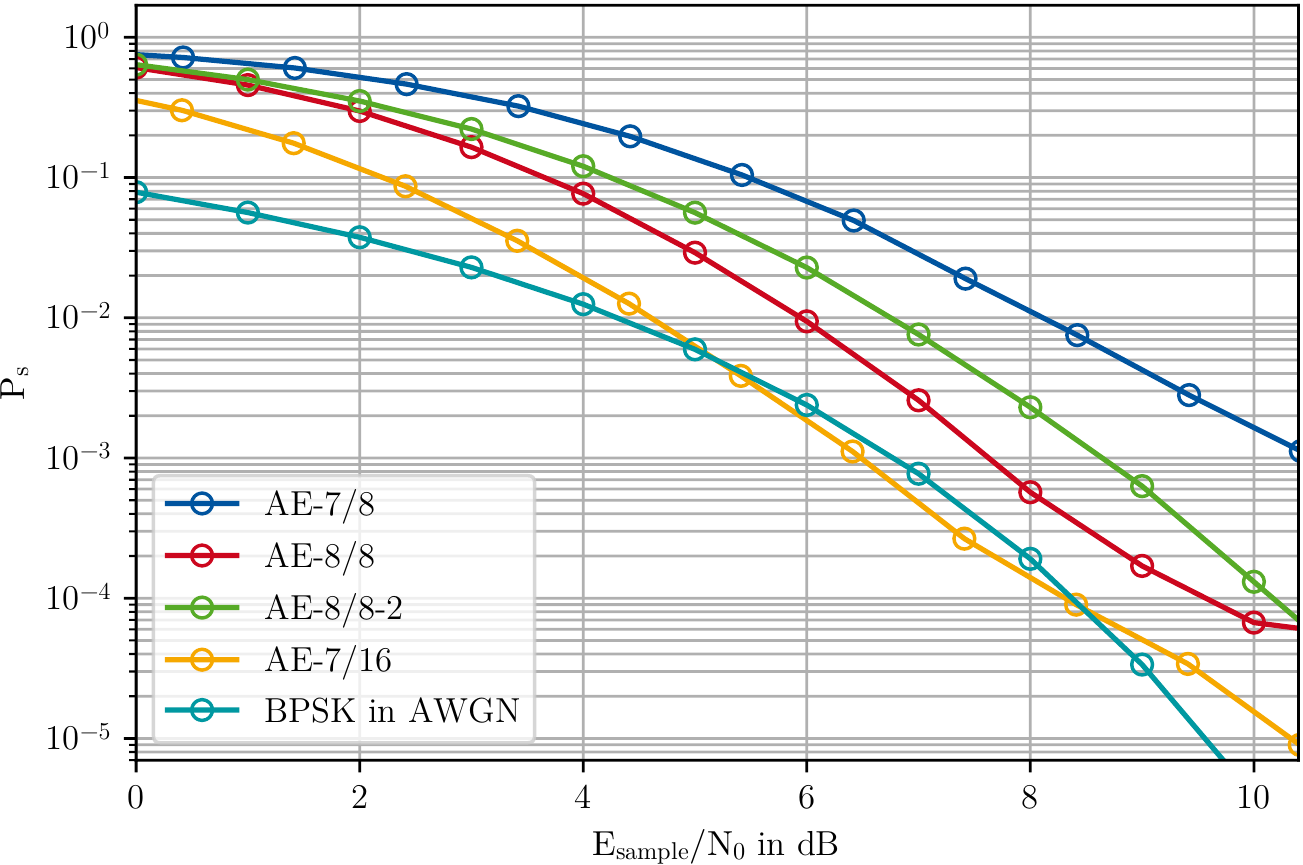}
		  \label{fig:ser_compare_Es}}
	\caption{SER of the four presented autoencoders, tested on $10^6$ symbols each over modeled channels, and theoretical BPSK error rate over AWGN as a baseline}
	\label{fig:ser_compare_model}
\end{figure}

In this section, the performance of the autoencoders is compared for the simulated channel. 
The channel attenuation is chosen as $a_{\text{min}}=0.01$. The phase shift is generated at random uniformly over $[0,2\pi)$ and the time offset chosen uniformly distributed over $\{-7,\dots,8\}$.   
$10^6$ data symbols are sent for each of the four autoencoders and for a range of different \acp{SNR}. We use two notions of \ac{SNR}, namely \ac{SNR} per sample $E_{\text{sample}}/N_0$ and  \ac{SNR} per bit $E_{\text{b}}/N_0$,

The \acp{SER} is plotted versus the \acp{SNR} per bit,  $E_{\text{b}}/N_0$ in Figure \ref{fig:ser_modelcompare}. As previously mentioned, the theoretical error rate of uncoded BPSK is plotted for comparison.
The performance of all autoencoders increases with \ac{SNR}. 
The AE-7/8 shows the slowest improvement over $E_{\text{b}}/N_0$ and reaches only an \acs{SER}  of $10^{-4}$ at 14\,dB.
While the AE-7/16 shows a similarly slow improvement for low \acp{SNR}, it improves more quickly and reaches the \ac{SER} below $10^{-5}$. Note that the AE-7/16, however, uses more complex samples than the AE-7/8 to transmit the same amount of data.

At a low \acp{SNR}, the best performance 
is achieved by the AE-8/8 and AE-8/8-2. The improvement of \ac{SER} with \ac{SNR} shows an error floor in the high \ac{SNR} regime. This is in contrast to the AE-7/16 which, at least within the tested range of \acp{SNR}, keeps improving and reaches a lower error rate than both the AE-8/8 and the AE-8/8-2. In that regard, the AE-7/16 outperforms other autoencoders in the high \ac{SNR} regime, however, at the price of using more samples to transmit data. 

We also plot the result of the same experiments for the \ac{SNR} per sample $E_{\text{sample}}/N_0$. The corresponding results for
the \acp{SER} are shown in Figure \ref{fig:ser_compare_Es}.
As expected, AE-7/16 performs better than other autoencoders, since it has more samples to transmit the same amount of data and can thus achieve a more robust representation of the encoded bits within the samples.
The AE-8/8 needs about 1.7\,dB more \ac{SNR} to achieve the same error rates as AE-7/16. This number is more than 2\,dB for AE-8/8-2.
AE-7/8 is significantly worse than the other autoencoders. 
If \ac{SNR} per sample is used, the AE-7/16 is able to outperform even uncoded BPSK for \ac{SNR} values between approximately 5 and 8\,dB. This can be attributed to the significantly greater number of complex samples used for transmission.

Uncoded BPSK mostly achieves better error rates than the proposed autoencoders. But as discussed above, this observation should be interpreted with care.
The theoretical BPSK error rate is determined for an AWGN channel, disregarding other effects of the channel model. Furthermore, the autoencoders are evaluated by their \acl{SER}, which is at most equal to the bit error rate, in most cases, however, significantly lower. Unlike the theoretical results for BPSK, the autoencoders of the present paper are designed to combat more channel impairments, as described above.

\begin{figure} 
	\centering
  \includegraphics[width=1\columnwidth]{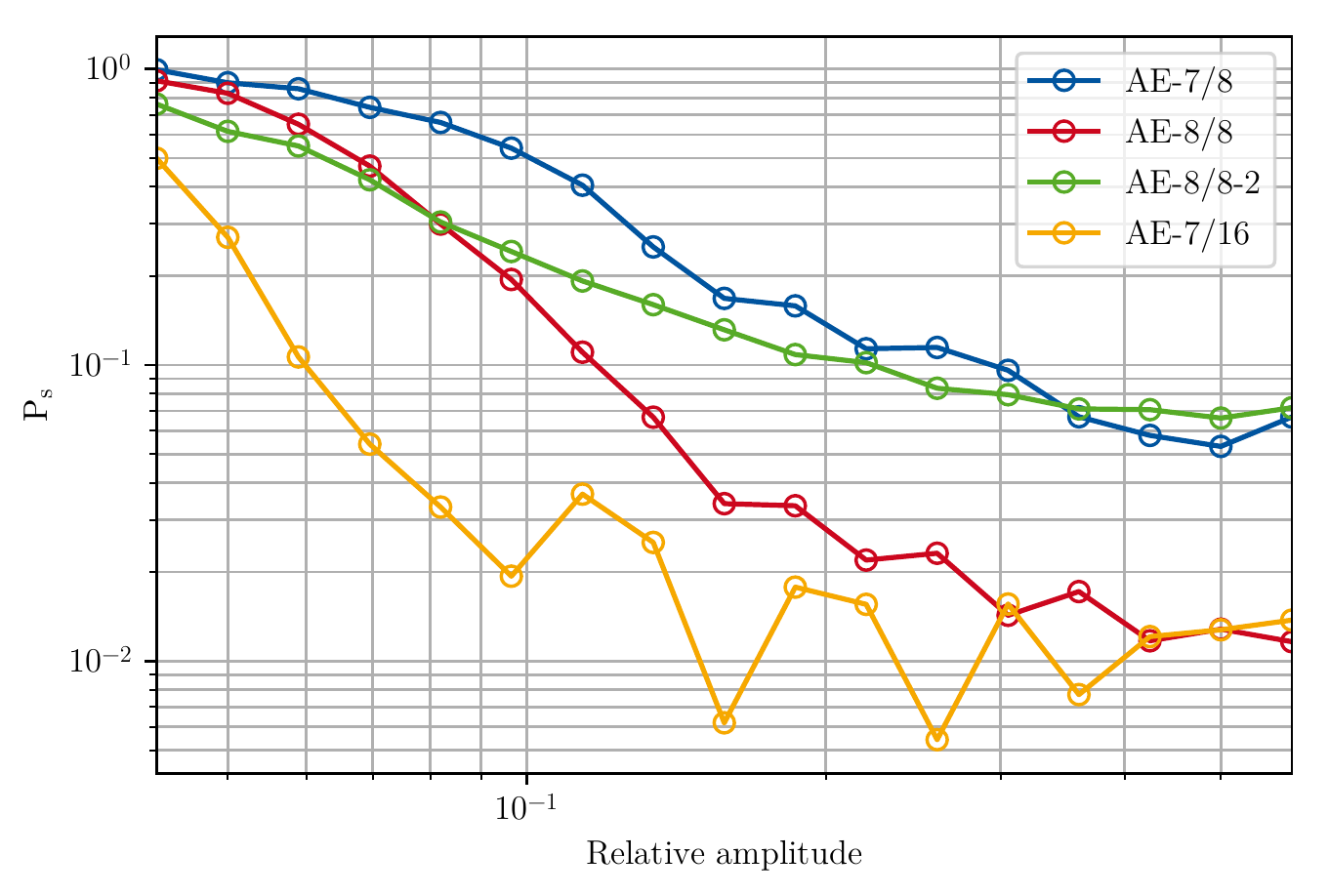}
  \caption{SER over the air for the four presented autoencoders, tested on $3.3\cdot10^6$ symbols each}
  \label{fig:ser_ota_unsyced}
\end{figure}

For over-the-air experiments, the \ac{SER} of the four autoencoders  is plotted in \ref{fig:ser_ota_unsyced}. We use $3.3\cdot10^6$ symbols for each autoencoder at different relative amplitude levels. Since the noise is added to each complex sample, the \ac{SNR} per sample is the natural choice for evaluating over-the-air transmission.

At low amplitudes, the AE-7/16 clearly outperforms the other autoencoders, and at amplitudes of about 0.2 it reaches the best error rates of all tested autoencoders with a value of about 0.6\%, see Fig.~\ref{fig:ser_unsynched}. Even for the worst \ac{SER} values in Figure \ref{fig:ser_unsynched}, the AE-7/16 shows better error rates over the whole range of tested amplitudes.
The next best performance is obtained by the AE-8/8 which reaches comparable performance only at high amplitudes. This is because its bit rate is more than two times bigger than AE-7/16's, and it thus has less redundancy to mitigate the effects of low amplitudes.
Interestingly, AE-8/8-2 performs better than AE-8/8 for low amplitudes, in contrast with what we observed in the simulations. On the other hand, the \ac{SER} of the AE-8/8-2 decreases very slowly and has an error floor at \ac{SER} 7\%, which is considerably worse than in the simulations.
Even the AE-7/8, which was consistently the worst autoencoder in our simulations, reaches lower error rates than the AE-8/8-3 while still showing inferior performance than AE-8/8 and AE-7/16.
When ranking the autoencoders with regard to their performance, the results of the over-the-air experiments coincide with the results of simulations, as shown in~\ref{fig:ser_compare_Es}.
Namely, the AE-7/16 performs better than the AE-8/8, followed by the AE-8/8-2 and the AE-7/8, which has the worst performance of the four.
It can be seen that the AE-8/8 demonstrates a surprisingly good over-the-air performance, considering its significantly higher data throughput than AE-7/16.

The bad performance of AE-8/8-2 emphasizes the importance of the \ac{SFE} block to improve decoding. The impact of the \ac{SFE} on the behavior of the autoencoders should be further investigated.
Whereas the AE-8/8-2, without the \ac{SFE}, performs  comparably well in the simulations, it is distinctly inferior to the  AE-8/8 over the air.

\newpage

\section{Conclusion and Outlook}
\label{sec:conclusion_outlook}
In this paper, we presented a fully trainable deep learning transceiver, which addresses in particular synchronization issues. Multiple autoencoders with different architectures are trained. Each has three different components, namely the encoder, the channel and the decoder. The  channel is incorporated in the training by using a model, thus enabling back propagation. The performance of the transceivers is evaluated by over the air transmission on an SDR platform. They achieve data rates of about 0.5 Mbps and contribute to high data rate deep learning transceivers. Future work will consider more realistic channel models, for example multi-path propagation models or frequency shifts, and training with different SNRs per batch. Recurrent neural networks can potentially learn tracking of synchronization parameters over time and can thereby improve the block based methods presented in this paper. Moreover, an autoencoder structure that transmits multiple data blocks per preamble block will be investigated to improve throughput.
\bibliographystyle{IEEEbib}
\bibliography{references}

\end{document}